\documentclass[aps,prb,nofootinbib,superscriptaddress,twocolumn]{revtex4}
\usepackage{graphicx}
\usepackage{epsfig}

\newcommand{\V}{V$_{15}$}
\newcommand{\Vfull}{K$_6$[V$^{IV}_{15}$As$_{6}$O$_{42}$(H$_2$O)]$\cdot8$H$_2$O}
\newcommand{\msr}{$\mu$SR}

\begin{document}

\title{Persistent Spin Dynamics in the $S=1/2$ V$_{15}$ Molecular Nano-Magnet}

\author{Z.~Salman}
\altaffiliation[Former address: ]{TRIUMF, 4004 Wesbrook Mall,
  Vancouver, BC, V6T 2A3, Canada}
\email{z.salman1@physics.ox.ac.uk}
\affiliation{Clarendon Laboratory, Department of
  Physics, Oxford University, Parks Road, Oxford OX1 3PU, UK}
\affiliation{ISIS Facility, Rutherford Appleton Laboratory, Chilton,
  Oxfordshire, OX11 0QX, UK}
\author{R.F.~Kiefl}
\affiliation{TRIUMF, 4004 Wesbrook Mall, Vancouver, BC, V6T 2A3, Canada}
\affiliation{Department of Physics and Astronomy, University of
  British Columbia, Vancouver, BC, V6T 1Z1, Canada}
\affiliation{Canadian Institute for Advanced Research, Canada}
\author{K.~H.~Chow}
\affiliation{Department of Physics, University of Alberta, Edmonton,
  AB, T6G 2G7, Canada}
\author{W.A.~MacFarlane}
\affiliation{Department of Chemistry, University of British Columbia, Vancouver, BC, V6T 1Z1, Canada}
\author{T.~Keeler}
\affiliation{Department of Physics and Astronomy, University of
  British Columbia, Vancouver, BC, V6T 1Z1, Canada}
\author{T.~Parolin}
\affiliation{Department of Chemistry, University of British Columbia, Vancouver, BC, V6T 1Z1, Canada}
\author{S.~Tabbara}
\affiliation{TRIUMF, 4004 Wesbrook Mall, Vancouver, BC, V6T 2A3, Canada}
\author{D.~Wang}
\affiliation{Department of Physics and Astronomy, University of
  British Columbia, Vancouver, BC, V6T 1Z1, Canada}

\begin{abstract}
We present muon spin lattice relaxation measurements in the \V\ spin
$1/2$ molecular nano-magnet. We find that the relaxation rate in low
magnetic fields ($<5$~kG) is temperature independent below $\sim
10$~K, implying that the molecular spin is dynamically fluctuating
down to $12$~mK. These measurements show that the fluctuation time
increases as the temperature is decreased and saturates at a value of
$\sim 6$ nsec at low temperatures. The fluctuations are attributed to
\V\ molecular spin dynamics perpendicular to the applied magnetic
field direction, induced by coupling between the molecular spin and
nuclear spin bath in the system.
\end{abstract}

\maketitle 
\section{Introduction}
Recent advances in the field of molecular magnetism, using a bottom up
approach to synthesis, has produced magnetic systems based on
molecules rather than metals or oxides. Example systems include single
molecule magnets (SMMs) and high spin molecules \cite{Sessoli03ACIE},
photo-magnets \cite{Sato96S,Sato99IC} and room temperature magnets
\cite{Ferlay95N}. The focus of the current work is SMMs and their
interaction with the environment. These systems have been attracting
much attention in recent years mainly due to the discovery of quantum
tunneling of the magnetization (QTM) in Mn$_{12}$ and Fe$_8$
\cite{Friedman96PRL,Thomas96N}. A SMM is composed of a small cluster
of magnetic ions with strong magnetic interactions between them ( $J
\sim 10-100$~K), embedded in magnetically inactive organic or
inorganic ligands. Each cluster is thus isolated from its neighbours,
forming at low temperatures, a lattice of non-interacting spins and
allowing the study of the quantum behaviour of isolated
spins. Molecular magnets are ideal objects to study phenomena of great
scientific importance for mesoscopic physics, such as QTM
\cite{Friedman96PRL,Thomas96N,Sangregorio97PRL}, topological quantum
phase interference \cite{Wernsdorfer99S,Wernsdorfer00EL}, and quantum
coherence
\cite{Dobrovitski00PRL,Barbara02PTPS,Hill03S,delBarco04PRL,Morello06PRL,Bertaina08N}. They
could also be applicable for the recording industry
\cite{Stamp96LTP,Joachim00N}, as well as for information transmission
and quantum computing
\cite{Divincenzo95QTM,Tejada01N,Troiani05PRL,Troiani05PRL2,Ardavan07PRL}.

The effective spin Hamiltonian describing a SMM is usually written as
\begin{equation} \label{Hamiltonian}
{\cal H} = - D S_z^2-g \mu_B {\mathbf S} \cdot {\mathbf H} + {\cal H}^t,
\end{equation}
where the first term is the easy axis magneto-crystalline anisotropy
term, which is the source of the double well structure and degeneracy
of spin up and spin down states, the second is the Zeeman term, and
${\cal H}^t$ includes additional crystal field terms as well as the
interaction between the molecular spin and its environment (including
molecule-molecule dipolar \cite{Prokofev98PRL,Morello06PRL} and
molecule-nuclear spin bath interactions
\cite{Prokofev98PRL,Keren07PRL}). In prototypical systems such as
Mn$_{12}$ and Fe$_{8}$, the anisotropy term is large and therefore
${\cal H}^t$ is only a perturbation, making it difficult to perform
direct measurements to identify the detailed nature of ${\cal
  H}^t$. However, isotropic ($D \simeq 0$) systems provide an ideal
system in which ${\cal H}^t$ can be probed directly. In this paper we
present muon spin relaxation (\msr) measurements on the spin 1/2
molecular magnet \Vfull\, known as \V\ \cite{Muller91ACIEE}. Due to
its spin 1/2 nature this molecule has no anisotropy, i.e. $D \equiv
0$. We find the molecular spin dynamics of \V\ persists down to mK
temperatures, and there is no evidence of freezing even in an applied
field of up to 10~kG.\cite{footnote2} The persistent spin dynamics is
consistent with a previously reported \msr\ study\cite{Procissi06PRB},
however, our measurements were extended to lower temperatures and
higher fields, allowing a more comprehensive understanding of the
nature and source of \V\ spin dynamics. The spin fluctuation time
measured is on the nanosecond scale and saturates at $\sim 6$ nsec at
low temperature. These molecular spin fluctuations are perpendicular
to the applied field direction and are attributed to the interaction
between the \V\ molecular spin and its environment, in particular
nuclear spins.

\begin{figure}[h]
\centerline{\includegraphics[height=5.0cm]{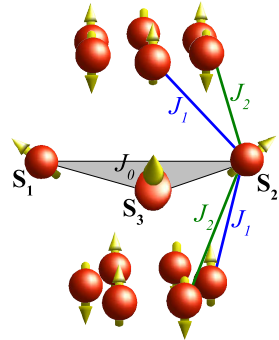}}
\caption{(Color online) The core of the \V\ molecule. Only V$^{IV}$
  ions are shown here. The arrows indicate the V$^{IV}$
  spins.}\label{V15core}
\end{figure}
The \V\ complex consists of a lattice of molecules with fifteen
V$^{IV}$ ions of spin $1/2$, in a quasi-spherical layered structure
formed of a triangle sandwiched between two hexagons (see
Fig.~\ref{V15core}). When the temperature is lower than $\sim 100$~K
the two hexagons of the V ions form a $S=0$ state due to
antiferromagnetic interactions, leaving the three V ions on the
triangle with an effective Hamiltonian
\cite{Chiorescu00PRL,Barbara02PTPS}
\begin{equation} \label{H0}
{\cal H}_0=-J_0 \left( {\mathbf S_1}\cdot {\mathbf S_2} + {\mathbf
  S_2} \cdot {\mathbf S_3} + {\mathbf S_3}\cdot {\mathbf S_1} \right)
  -g\mu_B {\mathbf H} \cdot \sum_{i=1}^3 {\mathbf S_i}
\end{equation}
where ${\mathbf S}_i$ are the spins of the three V ions and $J_0
\simeq -2.445$~K is the effective coupling between the spins, which is
due to competing $J_1$ and $J_2$ interactions (see
Fig. \ref{V15core}). At temperatures lower than $500$~mK the ground
state of each \V\ molecule is $S=1/2$, with negligible dipolar
interactions (few mK) between neighbouring molecules. In contrast to
Mn$_{12}$ and Fe$_8$, the \V\ molecule has no anisotropy barrier and a
large tunneling splitting at zero field, $\Delta_0 \simeq 80$~mK
\cite{Barbara02PTPS,Chiorescu03PRB}.

\section{Experiment}
The \msr\ experiments were performed on the M15 and M20 beamlines at
TRIUMF in Vancouver, Canada. In these experiments $100 \%$ polarized
(along $z$) positive muons are implanted in the sample. Each implanted
muon decays (lifetime $\tau=2.2$ $\mu$sec) emitting a positron
preferentially in the direction of its polarization at the time of
decay. Using appropriately positioned detectors, one measures the
asymmetry of the muon beta decay along $z$ as a function of time
$A(t)$, which is proportional to the time evolution of the muon spin
polarization. $A(t)$ depends on the distribution of internal magnetic
fields and their temporal fluctuations. Further details on the $\mu$SR
technique may be found in Ref.~\cite{Kilcoyne98,Chow98}.

\begin{figure}[h]
\centerline{\includegraphics[width=\columnwidth]{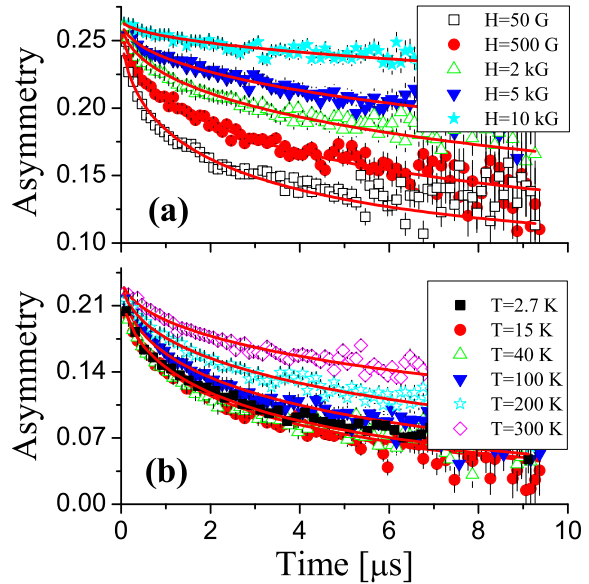}}
\caption{(Color online) Example muon spin relaxation curves at (a)
  $T=15$ mK and different magnetic fields, and (b) $H=50$ G and
  different temperatures. The solid lines are fits to
  Eq.~(\ref{rlx}).}\label{Asy}
\end{figure}
The composition and sturcture of \V\ single crystals were confirmed
using x-ray diffraction and their magnetization was examined in a
SQUID magnetometer. These single crystals were then crushed into a
fine powder and used for the \msr\ measurements reported here. The
fine powder was placed in a $^4$He gas flow cryostat to measure the
muon spin relaxation in the temperature range between 2.5 to 300
K. For the low temperature measurements the sample was pressed into
pellets and placed in a dilution refrigerator (DR) to measure the
relaxation between 12 mK and 3.5 K. To ensure thermalization of the
sample in the DR, the pellets were mounted on a silver plate using
Apiezon grease, wrapped in a silver foil and attached to the cold
finger of the DR. The relaxation in the full temperature range was
measured in magnetic fields up to 10 kG applied in the direction of
the initial muon spin polarization.

\section{Results}
\begin{figure}[h]
\centerline{\includegraphics[width=\columnwidth,clip]{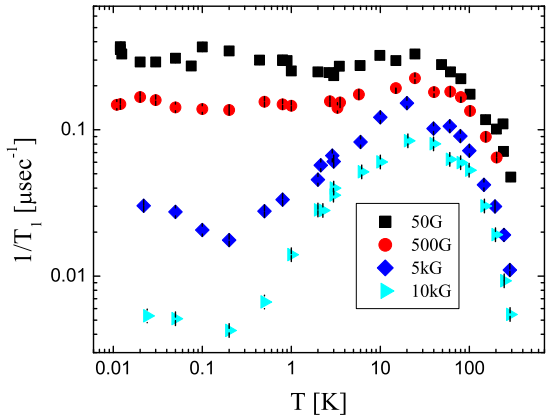}}
\caption{(Color online) The muon spin lattice relaxation rate as a
function of temperature for different values of external magnetic
field.} \label{T1vsTH}
\end{figure}

The measured asymmetry at all temperatures and applied fields was
found to fit best to a square root exponential function (see
Fig.~\ref{Asy})
\begin{equation} \label{rlx}
A(t)=A_0 e^{-\sqrt{\lambda t}}+A_{Bg},
\end{equation}
which reflects the dynamic nature of the local magnetic fields
experienced by the muons as well as multiple muon stopping sites
\cite{Salman02PRB,Uemura85PRB}. Here $A_0$ is the initial asymmetry
and $\lambda=1/T_1$ is the relaxation rate averaged over all stopping
sites.\cite{footnote1} $A_{Bg}$ is the non-relaxing background signal
due to muons missing the sample and stopping in the silver sample
holder. The high temperature measurements were performed using the low
background setup \cite{Schneider93PRL}. Therefore, a non-zero
background signal was found only in the measurements done in the
DR. The values of $\lambda$ in the full temperature range and various
magnetic fields are presented in Fig.~\ref{T1vsTH}.
The relaxation rate increases as the temperature is decreased from
room temperature down to $\sim 50$K due to the slowing down of the
thermally activated transitions between the different spin states
\cite{Salman02PRB,Lascialfari98PRL,Blundell04JPCM}. This increase in
the relaxation rate is a signature of the increased correlations
between the V moments, as the temperature becomes comparable to the
magnetic couplings inside the \V\ core. Similar behavior is observed
in many molecular magnets studied by $\mu$SR and NMR
\cite{Salman02PRB,Lascialfari98PRL,Blundell04JPCM,Borsa06,Borsa07}. Most
striking is the fact that the relaxation rate at temperatures below
$\sim 10$~K and at fields lower than $5$~kG becomes almost temperature
independent, in agreement with similar $^1$H
NMR\cite{Yoneda03PB,Furukawa05P,Procissi06PRB} and \msr
\cite{Procissi06PRB} studies. At higher fields ($5$ and $10$~kG) and
temperatures lower than $10$~K the relaxation rate decreases as the
temperature is decreased down to $200$~mK. Since at these temperatures
the $S=1/2$ ground spin state is preferentially populated, we
attribute the decrease in $1/T_1$ to the reduction in the thermally
activated transitions between the (Zeeman) split $m=\pm 1/2$ states,
in agreement with $^1$H NMR measurements\cite{Yoneda03PB,Furukawa05P}.
However, at even lower temperatures (less than $\sim 200$~mK) $1/T_1$
seems to saturate (within experimental accuracy) and become
temperature independent even at high fields. Note that $1/T_1$
exhibits a strong field dependence, which can be clearly seen in
Fig.~\ref{Asy}(a). In contrast, $1/T_1$ seems to be field independent
up to $1.5$ kG in Ref.~\onlinecite{Procissi06PRB}. The source of
discrepancy between these results and our results is still unclear.

A better understanding of the relaxation rate can be obtained by
studying the eigenstates and eigenvalues of the Hamiltonian
(\ref{H0}). We rewrite the Hamiltonian as
\begin{equation} \label{H0diag}
{\cal H}_0=-\frac{J_0}{2} \left( {\mathbf S^2}-\frac{9}{4} \right)
-g\mu_B {\mathbf H} \cdot {\mathbf S}
\end{equation}
where ${\mathbf S}=\sum_{i=1}^3 {\mathbf S}_i$ and $S$ are the total
spin operator and spin number, respectively. The eigenstates of ${\cal
H}_0$ are $|Sm\rangle$ with eigenvalues $E_{Sm}$, where $S$ can be
either $1/2$ or $3/2$ with $m=-S,-S+1,\cdots,S$. Note that the $S=1/2$
state is doubly degenerate. The energy level scheme implies that at
temperatures $T \ll -3 J_0/ 2 \simeq 3.7$~K (i.e. the splitting
between the $S=3/2$ and $1/2$ states), only the ground spin state
$S=1/2$ is appreciably populated. With the absence of additional terms
in ${\cal H}_0$, the molecular spin and hence the resultant local
field experienced by the muon should be static. However, as seen in
Fig.~\ref{T1vsTH} at low fields the relaxation rate is finite at these
temperatures, and therefore the local field (and molecular spin)
continues to fluctuate down to $12$~mK, the lowest temperature
measured. These results indicate that ${\cal H}_0$ cannot fully
describe the system at low temperature. Rather, one expects the
existence of an additional term in the spin Hamiltonian, ${\cal
H}^{t}$, that mixes the unperturbed $|Sm \rangle$ eigenstates, hence
inducing transitions between them.

We expect that the local magnetic field experienced by the implanted
muons is proportional to the \V\ neighbouring spins. Hence, assuming
that fluctuations of the \V\ spins, described by exponential
correlation functions \cite{Lowe68PR}, are
\begin{eqnarray}
\left\langle S_z(t)S_z(0) \right\rangle &=& \frac{1}{3} S(S+1)
\exp\left(-\frac{t}{\tau_z}\right) \label{correlationz}\\
\left\langle S_{+}(t)S_{-}(0) \right\rangle &=& \frac{2}{3} S(S+1)
\exp(i \omega_e t) \exp\left(-\frac{t}{\tau_{\pm}}\right) \label{correlationxy},
\end{eqnarray}
where $\tau_z$ and $\tau_{\pm}$ are the correlation times of the
parallel and perpendicular (to the applied field) ${\mathbf S}$
components respectively, and $\omega_e=g \mu_B H$ is the electronic
resonance frequency. For simplicity we assume that the correlation
times for both components are equal, $\tau_z=\tau_{\pm}=\tau$. In this
case $1/T_1$ can be written as \cite{Lowe68PR}
\begin{equation} \label{SLRexact}
\frac{1}{T_1}=\frac{2 \Delta_z^2 \tau}{1+\omega_\mu^2
  \tau^2}+\frac{2\Delta_+^2 \tau}{1+(\omega_e+\omega_\mu)^2\tau^2}
  +\frac{2\Delta_-^2 \tau}{1+(\omega_e-\omega_\mu)^2\tau^2},
\end{equation}
where $\omega_\mu=\gamma H$ is the muon Larmor frequency and
$\Delta_i$ is an estimate of the coupling strength between the muon
and the molecular spin, which depends on the nature of the coupling
(e.g. exchange or dipolar). The first term in Eq.~(\ref{SLRexact})
originates from flipping of the muon while second and third terms are
due to flip-flop and co-flipping of the muon and molecular spins,
respectively. Using $\omega_e \gg \omega_{\mu}$ one can rewrite
Eq.~(\ref{SLRexact}) as \cite{Dunsiger06PRB},
\begin{equation} \label{SLR}
\frac{1}{T_1} \approx 2 f \frac{\Delta^2 \tau}{1+\omega_\mu^2
  \tau^2}+2(1-f)\frac{\Delta^2 \tau}{1+\omega_e^2 \tau^2}
\end{equation}
where $\Delta=\sqrt{\Delta_z^2+\Delta_+^2+\Delta_-^2}$ is a measure of
the distribution of local magnetic field observed by muons, which
depends on the coupling strength between the muon spin ${\mathbf I}$
and molecular spin ${\mathbf S}$, and $f=\Delta_z^2 / \Delta^2$. Note
the first term in Eq.~(\ref{SLR}) is due to fluctuations along the
applied field ($S_z$), while the second is due to fluctuations in
$S_x$ and $S_y$. The value of $f$ measures the relative size of
parallel versus perpendicular local field. For example, using the
assumptions above, for a dipolar interaction between ${\mathbf I}$ and
${\mathbf S}$ (which is fluctuating randomly in all directions) we
expect on average $f \approx 0.3$. However, if our earlier assumption
of $\tau_z=\tau_{\pm}$ in Eqs.~(\ref{correlationz}) and
(\ref{correlationxy}) does not hold, $f$ will contain also information
on the importance of parallel versus perpendicular fluctuations,
e.g. $f=0$ implies that $\tau_{z} \gg \tau_{\pm}$ and therefore
$1/T_1$ is dominated by fluctuations perpendicular to the applied
field.
\begin{figure}[h]
\centerline{\includegraphics[width=\columnwidth,clip]{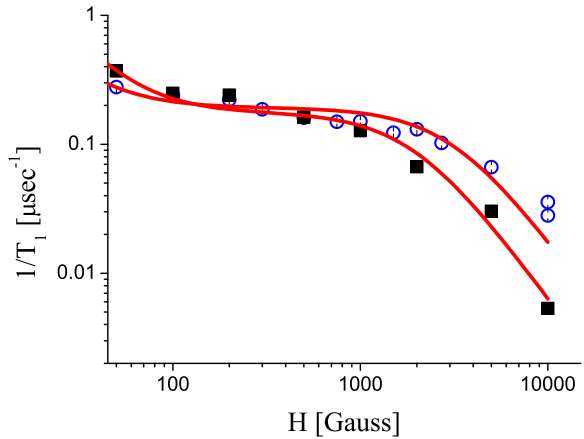}}
\caption{(Color online) The spin lattice relaxation rate as a function
  of the applied magnetic field for $T=15$~mK (squares) and $T=2.7$~K
  (circles). The solid lines are fits to Eq.~(\ref{SLR}).}
  \label{T1vsH}
\end{figure} 

\begin{figure}[h]
\centerline{\includegraphics[width=\columnwidth,clip]{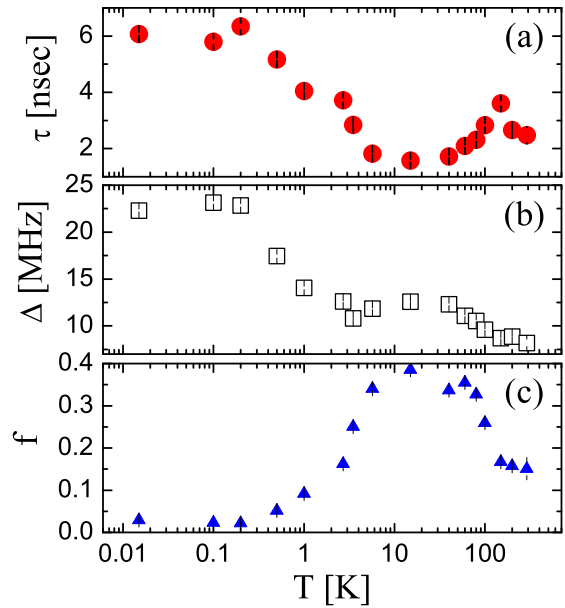}}
\caption{(Color online) The correlation time $\tau$ (a), the size of
  the local field experienced by muons $\Delta$ (b), and the value of
  $f$ (c) as a function of temperature.}
  \label{taufDel}
\end{figure}

Next, we evaluate the correlation time $\tau$ from the field
dependence of $1/T_1$ using Eq.~(\ref{SLR}). Note that $\tau$ is the
fluctuation time of the molecular spin. In Fig.~\ref{T1vsH} we present
the relaxation rate as a function of applied magnetic field for two
different temperatures. The solid lines are fits of the data to
Eq.~(\ref{SLR}), from which the values of $\tau$, $\Delta$ and $f$ for
the corresponding temperature are extracted. By fitting the field
dependence of $1/T_1$ for all temperatures we obtain the values of
these parameters as a function of temperature, shown in
Fig.~\ref{taufDel}. In Fig.~\ref{taufDel}(a) the correlation time
increases from $\tau \sim 1-2$~nsec above $T=10$~K and saturates at
$\tau \sim 6$~nsec below $1$~K. The value of this correlation time is
similar to that measured earlier in other isotropic high spin molecule
systems \cite{Salman00PB,Salman02PRB,Keren07PRL}. This similarity
strongly suggests that a similar mechanism is responsible for the
observed molecular spin fluctuations, namely an interaction between
the molecular spin and its environment, in particular the interaction
with neighbouring nuclear moments (nuclear spin bath), inducing
transitions between the different states
\cite{Salman02PRB,Keren07PRL}. In Ref.~\onlinecite{Procissi06PRB} the
low temperature dynamics were attributed to spin fluctuations between
the two lowest lying $S=1/2$ doublet, whose energy difference is $80$
mK. However, our measurement show that the dynamics persist down to 12
mK, much lower than the splitting of the doublet. Therefore the
proposed source of fluctuations in Ref. \cite{Procissi06PRB} cannot
account for the dynamics observed in \V.

The value of $\Delta$, shown in Fig.~\ref{taufDel}(b), increases as
the temperature is decreased from room temperature down to
$\sim100$~K. This increase coincides with the observed decrease in the
effective magnetic moment above
$100$~K\cite{Barbara02PTPS}. Similarly, the increase below $\sim 1$~K
coincides with the additional decrease in the effective magnetic
moment from $3$~$\mu_B$ down to $1$~$\mu_B$ at this temperature. Note
that $\Delta$ is a measure of the distribution of the local magnetic
fields experienced by muons, and therefore is not simply proportional
to the measured \V\ magnetic moment. Instead, factors such as the
local magnetization combined with the fluctuation rate as well as the
type of interaction between the \V\ moments and muons determine the
value of $\Delta$.

\section{Summary and Conclusions}
As we mentioned earlier, if the parallel and perpendicular
fluctuations are different, the value of $f$ is related to the
direction of fluctuations of ${\mathbf S}$, perpendicular or parallel
to the magnetic field (defined as $z$ direction). Interestingly, we
find $f \sim 0.35$ at high temperatures ($5-100$ K) (see
Fig.~\ref{taufDel}(c)), indicating that (I) the fluctuations of
${\mathbf S}$ are in randomly oriented relative to the applied field
as expected from thermal fluctuations and (II) the coupling between
${\mathbf I}$ and ${\mathbf S}$ is dipolar. As the temperature is
reduced, thermal fluctuations between the spin states become less
likely, especially when the temperature is much lower than the energy
splitting between the levels (including Zeeman). Nevertheless, the
fluctuations persist even at mK temperatures. Note however, that at
low temperatures $f \approx 0$, i.e. the fluctuations are
predominantly perpendicular to the $z$ direction and do not originate
from transitions between, e.g. $m=\pm 1/2$ spin states, since these
are indeed separated by the Zeeman energy. Instead they are most
probably due to the finite energy level broadening produced by the
interaction between ${\mathbf S}$ and the nuclear spin bath in the
system. This occurs since in the presence of the nuclear moments (as a
perturbation on ${\cal H}_0$), $S_z$ is no longer a good quantum
number, and therefore the $m$ levels are split or broadened. The size
of $\tau \sim 6$ nsec implies broadening of the levels on the scale of
few tens of mK, which is a reasonable size for such interactions
\cite{Salman02PRB,Keren07PRL}. Finally, since at $T>100$ K there are
less correlations between the V ions in the magnetic core, the
fluctuations measured by the muon are due to its interaction with
multiple, uncorrelated V ions. This may be the reason for the decrease
of $f$ at $T>100$ K. Alternatively, it may also be due to diffusion of
the muons in the sample at these temperatures.

In conclusion, we found that the muon spin lattice relaxation rate is
temperature independent at low temperatures and fields, in agreement
with earlier NMR and
\msr\ measurements.\cite{Yoneda03PB,Furukawa05P,Procissi06PRB} This
temperature independence exhibits the non-thermal/quantum nature of
the \V\ molecular spin fluctuations which persist down to $12$ mK
temperature. The correlation time of the spin-spin auto-correlation
functions was estimated to be $\sim 6$ nsec at low temperature. We
believe that this low temperature spin dynamics is due to broadening
of the spin states introduced by the interactions between the \V\ spin
and the nuclear spin bath.

\begin{acknowledgments}
We would like to thank Achim M\"uller at the University of Bielefeld
for providing the sample and Bernard Barbara at Institut N\'eel for
helpful discussions. This work was supported by the CIAR, NSERC and
TRIUMF. We thank Syd Kreitzman, Rahim Abasalti, Bassam Hitti and
Donald Arseneau for technical support.
\end{acknowledgments}

\newcommand{\noopsort}[1]{} \newcommand{\printfirst}[2]{#1}
  \newcommand{\singleletter}[1]{#1} \newcommand{\switchargs}[2]{#2#1}


\begin{thebibliography}{42}
\expandafter\ifx\csname natexlab\endcsname\relax\def\natexlab#1{#1}\fi
\expandafter\ifx\csname bibnamefont\endcsname\relax
  \def\bibnamefont#1{#1}\fi
\expandafter\ifx\csname bibfnamefont\endcsname\relax
  \def\bibfnamefont#1{#1}\fi
\expandafter\ifx\csname citenamefont\endcsname\relax
  \def\citenamefont#1{#1}\fi
\expandafter\ifx\csname url\endcsname\relax
  \def\url#1{\texttt{#1}}\fi
\expandafter\ifx\csname urlprefix\endcsname\relax\def\urlprefix{URL }\fi
\providecommand{\bibinfo}[2]{#2}
\providecommand{\eprint}[2][]{\url{#2}}

\bibitem[{\citenamefont{Gatteschi and Sessoli}(2003)}]{Sessoli03ACIE}
\bibinfo{author}{\bibfnamefont{D.}~\bibnamefont{Gatteschi}} \bibnamefont{and}
  \bibinfo{author}{\bibfnamefont{R.}~\bibnamefont{Sessoli}},
  \bibinfo{journal}{Angew. Chem. Int. Ed.} \textbf{\bibinfo{volume}{42}},
  \bibinfo{pages}{268} (\bibinfo{year}{2003}).

\bibitem[{\citenamefont{Sato et~al.}(1996)\citenamefont{Sato, Iyoda, Fujishima,
  and Hashimoto}}]{Sato96S}
\bibinfo{author}{\bibfnamefont{O.}~\bibnamefont{Sato}},
  \bibinfo{author}{\bibfnamefont{T.}~\bibnamefont{Iyoda}},
  \bibinfo{author}{\bibfnamefont{A.}~\bibnamefont{Fujishima}},
  \bibnamefont{and}
  \bibinfo{author}{\bibfnamefont{K.}~\bibnamefont{Hashimoto}},
  \bibinfo{journal}{Science} \textbf{\bibinfo{volume}{272}},
  \bibinfo{pages}{704} (\bibinfo{year}{1996}).

\bibitem[{\citenamefont{Sato et~al.}(1999)\citenamefont{Sato, Einaga,
  Fujishima, and Hashimoto}}]{Sato99IC}
\bibinfo{author}{\bibfnamefont{O.}~\bibnamefont{Sato}},
  \bibinfo{author}{\bibfnamefont{Y.}~\bibnamefont{Einaga}},
  \bibinfo{author}{\bibfnamefont{A.}~\bibnamefont{Fujishima}},
  \bibnamefont{and}
  \bibinfo{author}{\bibfnamefont{K.}~\bibnamefont{Hashimoto}},
  \bibinfo{journal}{Inorg. Chem.} \textbf{\bibinfo{volume}{38}},
  \bibinfo{pages}{4405} (\bibinfo{year}{1999}).

\bibitem[{\citenamefont{Ferlay et~al.}(1995)\citenamefont{Ferlay, Mallah,
  Ouahes, Villet, and Verdaguer}}]{Ferlay95N}
\bibinfo{author}{\bibfnamefont{S.}~\bibnamefont{Ferlay}},
  \bibinfo{author}{\bibfnamefont{T.}~\bibnamefont{Mallah}},
  \bibinfo{author}{\bibfnamefont{R.}~\bibnamefont{Ouahes}},
  \bibinfo{author}{\bibfnamefont{P.}~\bibnamefont{Villet}}, \bibnamefont{and}
  \bibinfo{author}{\bibfnamefont{M.}~\bibnamefont{Verdaguer}},
  \bibinfo{journal}{Nature} \textbf{\bibinfo{volume}{378}},
  \bibinfo{pages}{701} (\bibinfo{year}{1995}).

\bibitem[{\citenamefont{Thomas et~al.}(1996)\citenamefont{Thomas, Lionti,
  Ballou, Gatteschi, Sessoli, and Barbara}}]{Thomas96N}
\bibinfo{author}{\bibfnamefont{L.}~\bibnamefont{Thomas}},
  \bibinfo{author}{\bibfnamefont{F.}~\bibnamefont{Lionti}},
  \bibinfo{author}{\bibfnamefont{R.}~\bibnamefont{Ballou}},
  \bibinfo{author}{\bibfnamefont{D.}~\bibnamefont{Gatteschi}},
  \bibinfo{author}{\bibfnamefont{R.}~\bibnamefont{Sessoli}}, \bibnamefont{and}
  \bibinfo{author}{\bibfnamefont{B.}~\bibnamefont{Barbara}},
  \bibinfo{journal}{Nature (London)} \textbf{\bibinfo{volume}{383}},
  \bibinfo{pages}{145} (\bibinfo{year}{1996}).

\bibitem[{\citenamefont{Friedman et~al.}(1996)\citenamefont{Friedman, Sarachik,
  Tejada, and Ziolo}}]{Friedman96PRL}
\bibinfo{author}{\bibfnamefont{J.~R.} \bibnamefont{Friedman}},
  \bibinfo{author}{\bibfnamefont{M.~P.} \bibnamefont{Sarachik}},
  \bibinfo{author}{\bibfnamefont{J.}~\bibnamefont{Tejada}}, \bibnamefont{and}
  \bibinfo{author}{\bibfnamefont{R.}~\bibnamefont{Ziolo}},
  \bibinfo{journal}{Phys. Rev. Lett.} \textbf{\bibinfo{volume}{76}},
  \bibinfo{pages}{3830} (\bibinfo{year}{1996}).

\bibitem[{\citenamefont{Sangregorio et~al.}(1997)\citenamefont{Sangregorio,
  Ohm, Paulsen, Sessoli, and Gatteschi}}]{Sangregorio97PRL}
\bibinfo{author}{\bibfnamefont{C.}~\bibnamefont{Sangregorio}},
  \bibinfo{author}{\bibfnamefont{T.}~\bibnamefont{Ohm}},
  \bibinfo{author}{\bibfnamefont{C.}~\bibnamefont{Paulsen}},
  \bibinfo{author}{\bibfnamefont{R.}~\bibnamefont{Sessoli}}, \bibnamefont{and}
  \bibinfo{author}{\bibfnamefont{D.}~\bibnamefont{Gatteschi}},
  \bibinfo{journal}{Phys. Rev. Lett.} \textbf{\bibinfo{volume}{78}},
  \bibinfo{pages}{4645} (\bibinfo{year}{1997}).

\bibitem[{\citenamefont{Wernsdorfer and Sessoli}(1999)}]{Wernsdorfer99S}
\bibinfo{author}{\bibfnamefont{W.}~\bibnamefont{Wernsdorfer}} \bibnamefont{and}
  \bibinfo{author}{\bibfnamefont{R.}~\bibnamefont{Sessoli}},
  \bibinfo{journal}{Science} \textbf{\bibinfo{volume}{284}},
  \bibinfo{pages}{133} (\bibinfo{year}{1999}).

\bibitem[{\citenamefont{Wernsdorfer et~al.}(2000)\citenamefont{Wernsdorfer,
  Sessoli, Caneschi, Gatteschi, and Cornia}}]{Wernsdorfer00EL}
\bibinfo{author}{\bibfnamefont{W.}~\bibnamefont{Wernsdorfer}},
  \bibinfo{author}{\bibfnamefont{R.}~\bibnamefont{Sessoli}},
  \bibinfo{author}{\bibfnamefont{A.}~\bibnamefont{Caneschi}},
  \bibinfo{author}{\bibfnamefont{D.}~\bibnamefont{Gatteschi}},
  \bibnamefont{and} \bibinfo{author}{\bibfnamefont{A.}~\bibnamefont{Cornia}},
  \bibinfo{journal}{Europhys. Lett.} \textbf{\bibinfo{volume}{50}},
  \bibinfo{pages}{552} (\bibinfo{year}{2000}).

\bibitem[{\citenamefont{Dobrovitski et~al.}(2000)\citenamefont{Dobrovitski,
  Katsnelson, and Harmon}}]{Dobrovitski00PRL}
\bibinfo{author}{\bibfnamefont{V.~V.} \bibnamefont{Dobrovitski}},
  \bibinfo{author}{\bibfnamefont{M.~I.} \bibnamefont{Katsnelson}},
  \bibnamefont{and} \bibinfo{author}{\bibfnamefont{B.~N.}
  \bibnamefont{Harmon}}, \bibinfo{journal}{Phys. Rev. Lett.}
  \textbf{\bibinfo{volume}{84}}, \bibinfo{pages}{3458} (\bibinfo{year}{2000}).

\bibitem[{\citenamefont{Barbara et~al.}(2002)\citenamefont{Barbara, Chiorescu,
  Wernsdorfer, B\"ogge, and M\"uller}}]{Barbara02PTPS}
\bibinfo{author}{\bibfnamefont{B.}~\bibnamefont{Barbara}},
  \bibinfo{author}{\bibfnamefont{I.}~\bibnamefont{Chiorescu}},
  \bibinfo{author}{\bibfnamefont{W.}~\bibnamefont{Wernsdorfer}},
  \bibinfo{author}{\bibfnamefont{H.}~\bibnamefont{B\"ogge}}, \bibnamefont{and}
  \bibinfo{author}{\bibfnamefont{A.}~\bibnamefont{M\"uller}},
  \bibinfo{journal}{Prog. Theor. Phys. Supp.} \textbf{\bibinfo{volume}{145}},
  \bibinfo{pages}{357} (\bibinfo{year}{2002}).

\bibitem[{\citenamefont{Morello et~al.}(2006)\citenamefont{Morello, Stamp, and
  Tupitsyn}}]{Morello06PRL}
\bibinfo{author}{\bibfnamefont{A.}~\bibnamefont{Morello}},
  \bibinfo{author}{\bibfnamefont{P.~C.~E.} \bibnamefont{Stamp}},
  \bibnamefont{and} \bibinfo{author}{\bibfnamefont{I.~S.}
  \bibnamefont{Tupitsyn}}, \bibinfo{journal}{Phys. Rev. Lett.}
  \textbf{\bibinfo{volume}{97}}, \bibinfo{pages}{207206}
  (\bibinfo{year}{2006}).

\bibitem[{\citenamefont{Hill et~al.}(2003)\citenamefont{Hill, Edwards,
  Aliaga-Alcalde, and Christou}}]{Hill03S}
\bibinfo{author}{\bibfnamefont{S.}~\bibnamefont{Hill}},
  \bibinfo{author}{\bibfnamefont{R.~S.} \bibnamefont{Edwards}},
  \bibinfo{author}{\bibfnamefont{N.}~\bibnamefont{Aliaga-Alcalde}},
  \bibnamefont{and} \bibinfo{author}{\bibfnamefont{G.}~\bibnamefont{Christou}},
  \bibinfo{journal}{Science} \textbf{\bibinfo{volume}{302}},
  \bibinfo{pages}{1015} (\bibinfo{year}{2003}).

\bibitem[{\citenamefont{del Barco et~al.}(2004)\citenamefont{del Barco, Kent,
  Yang, and Hendrickson}}]{delBarco04PRL}
\bibinfo{author}{\bibfnamefont{E.}~\bibnamefont{del Barco}},
  \bibinfo{author}{\bibfnamefont{A.~D.} \bibnamefont{Kent}},
  \bibinfo{author}{\bibfnamefont{E.~C.} \bibnamefont{Yang}}, \bibnamefont{and}
  \bibinfo{author}{\bibfnamefont{D.~N.} \bibnamefont{Hendrickson}},
  \bibinfo{journal}{Phys. Rev. Lett.} \textbf{\bibinfo{volume}{93}},
  \bibinfo{pages}{157202} (\bibinfo{year}{2004}).

\bibitem[{\citenamefont{Bertaina et~al.}(2008)\citenamefont{Bertaina,
  Gambarelli, Mitra, Tsukerblat, M\"uller, and Barbara}}]{Bertaina08N}
\bibinfo{author}{\bibfnamefont{S.}~\bibnamefont{Bertaina}},
  \bibinfo{author}{\bibfnamefont{S.}~\bibnamefont{Gambarelli}},
  \bibinfo{author}{\bibfnamefont{T.}~\bibnamefont{Mitra}},
  \bibinfo{author}{\bibfnamefont{B.}~\bibnamefont{Tsukerblat}},
  \bibinfo{author}{\bibfnamefont{A.}~\bibnamefont{M\"uller}}, \bibnamefont{and}
  \bibinfo{author}{\bibfnamefont{B.}~\bibnamefont{Barbara}},
  \bibinfo{journal}{Nature} \textbf{\bibinfo{volume}{453}},
  \bibinfo{pages}{203} (\bibinfo{year}{2008}).

\bibitem[{\citenamefont{Prokof\'ev and Stamp}(1996)}]{Stamp96LTP}
\bibinfo{author}{\bibfnamefont{N.~V.} \bibnamefont{Prokof\'ev}}
  \bibnamefont{and} \bibinfo{author}{\bibfnamefont{P.~C.~E.}
  \bibnamefont{Stamp}}, \bibinfo{journal}{J. Low Temp. Phys.}
  \textbf{\bibinfo{volume}{104}}, \bibinfo{pages}{143} (\bibinfo{year}{1996}).

\bibitem[{\citenamefont{Joachim et~al.}(2000)\citenamefont{Joachim, Gimzewski,
  and Aviram}}]{Joachim00N}
\bibinfo{author}{\bibfnamefont{C.}~\bibnamefont{Joachim}},
  \bibinfo{author}{\bibfnamefont{J.~K.} \bibnamefont{Gimzewski}},
  \bibnamefont{and} \bibinfo{author}{\bibfnamefont{A.}~\bibnamefont{Aviram}},
  \bibinfo{journal}{Nature} \textbf{\bibinfo{volume}{408}},
  \bibinfo{pages}{541} (\bibinfo{year}{2000}).

\bibitem[{\citenamefont{Divincenzo}(1995)}]{Divincenzo95QTM}
\bibinfo{author}{\bibfnamefont{D.}~\bibnamefont{Divincenzo}}, in
  \emph{\bibinfo{booktitle}{Quantum Tunneling of the Magnetization - QTM '94}},
  edited by \bibinfo{editor}{\bibfnamefont{L.}~\bibnamefont{Gunther}}
  \bibnamefont{and} \bibinfo{editor}{\bibfnamefont{B.}~\bibnamefont{Barbara}}
  (\bibinfo{publisher}{Kluwer Publishing, Dordrecht}, \bibinfo{year}{1995}), p.
  \bibinfo{pages}{189}.

\bibitem[{\citenamefont{Tejada et~al.}(2001)\citenamefont{Tejada, Chudnovsky,
  del Barco, Hernandez, and Spiller}}]{Tejada01N}
\bibinfo{author}{\bibfnamefont{J.}~\bibnamefont{Tejada}},
  \bibinfo{author}{\bibfnamefont{E.~M.} \bibnamefont{Chudnovsky}},
  \bibinfo{author}{\bibfnamefont{E.}~\bibnamefont{del Barco}},
  \bibinfo{author}{\bibfnamefont{J.~M.} \bibnamefont{Hernandez}},
  \bibnamefont{and} \bibinfo{author}{\bibfnamefont{T.~P.}
  \bibnamefont{Spiller}}, \bibinfo{journal}{Nanotechnology}
  \textbf{\bibinfo{volume}{12}}, \bibinfo{pages}{181} (\bibinfo{year}{2001}).

\bibitem[{\citenamefont{Troiani
  et~al.}(2005{\natexlab{a}})\citenamefont{Troiani, Affronte, Carretta,
  Santini, and Amoretti}}]{Troiani05PRL}
\bibinfo{author}{\bibfnamefont{F.}~\bibnamefont{Troiani}},
  \bibinfo{author}{\bibfnamefont{M.}~\bibnamefont{Affronte}},
  \bibinfo{author}{\bibfnamefont{S.}~\bibnamefont{Carretta}},
  \bibinfo{author}{\bibfnamefont{P.}~\bibnamefont{Santini}}, \bibnamefont{and}
  \bibinfo{author}{\bibfnamefont{G.}~\bibnamefont{Amoretti}},
  \bibinfo{journal}{Phys. Rev. Lett.} \textbf{\bibinfo{volume}{94}},
  \bibinfo{pages}{190501} (\bibinfo{year}{2005}{\natexlab{a}}).

\bibitem[{\citenamefont{Troiani
  et~al.}(2005{\natexlab{b}})\citenamefont{Troiani, Ghirri, Affronte, Carretta,
  Santini, Amoretti, Piligkos, Timco, and Winpenny}}]{Troiani05PRL2}
\bibinfo{author}{\bibfnamefont{F.}~\bibnamefont{Troiani}},
  \bibinfo{author}{\bibfnamefont{A.}~\bibnamefont{Ghirri}},
  \bibinfo{author}{\bibfnamefont{M.}~\bibnamefont{Affronte}},
  \bibinfo{author}{\bibfnamefont{S.}~\bibnamefont{Carretta}},
  \bibinfo{author}{\bibfnamefont{P.}~\bibnamefont{Santini}},
  \bibinfo{author}{\bibfnamefont{G.}~\bibnamefont{Amoretti}},
  \bibinfo{author}{\bibfnamefont{S.}~\bibnamefont{Piligkos}},
  \bibinfo{author}{\bibfnamefont{G.}~\bibnamefont{Timco}}, \bibnamefont{and}
  \bibinfo{author}{\bibfnamefont{R.~E.~P.} \bibnamefont{Winpenny}},
  \bibinfo{journal}{Phys. Rev. Lett.} \textbf{\bibinfo{volume}{94}},
  \bibinfo{pages}{207208} (\bibinfo{year}{2005}{\natexlab{b}}).

\bibitem[{\citenamefont{Ardavan et~al.}(2007)\citenamefont{Ardavan, Rival,
  Morton, Blundell, Tyryshkin, Timco, and Winpenny}}]{Ardavan07PRL}
\bibinfo{author}{\bibfnamefont{A.}~\bibnamefont{Ardavan}},
  \bibinfo{author}{\bibfnamefont{O.}~\bibnamefont{Rival}},
  \bibinfo{author}{\bibfnamefont{J.~J.~L.} \bibnamefont{Morton}},
  \bibinfo{author}{\bibfnamefont{S.~J.} \bibnamefont{Blundell}},
  \bibinfo{author}{\bibfnamefont{A.~M.} \bibnamefont{Tyryshkin}},
  \bibinfo{author}{\bibfnamefont{G.~A.} \bibnamefont{Timco}}, \bibnamefont{and}
  \bibinfo{author}{\bibfnamefont{R.~E.~P.} \bibnamefont{Winpenny}},
  \bibinfo{journal}{Phys. Rev. Lett.} \textbf{\bibinfo{volume}{98}},
  \bibinfo{pages}{057201} (\bibinfo{year}{2007}).

\bibitem[{\citenamefont{Prokof\'ev and Stamp}(1998)}]{Prokofev98PRL}
\bibinfo{author}{\bibfnamefont{N.~V.}~\bibnamefont{Prokof\'ev}} \bibnamefont{and}
  \bibinfo{author}{\bibfnamefont{P.~C.~E.} \bibnamefont{Stamp}},
  \bibinfo{journal}{Phys. Rev. Lett.} \textbf{\bibinfo{volume}{80}},
  \bibinfo{pages}{5794} (\bibinfo{year}{1998}).

\bibitem[{\citenamefont{Keren et~al.}(2007)\citenamefont{Keren, Shafir,
  Shimshoni, Marvaud, Bachschmidt, and Long}}]{Keren07PRL}
\bibinfo{author}{\bibfnamefont{A.}~\bibnamefont{Keren}},
  \bibinfo{author}{\bibfnamefont{O.}~\bibnamefont{Shafir}},
  \bibinfo{author}{\bibfnamefont{E.}~\bibnamefont{Shimshoni}},
  \bibinfo{author}{\bibfnamefont{V.}~\bibnamefont{Marvaud}},
  \bibinfo{author}{\bibfnamefont{A.}~\bibnamefont{Bachschmidt}},
  \bibnamefont{and} \bibinfo{author}{\bibfnamefont{J.}~\bibnamefont{Long}},
  \bibinfo{journal}{Phys. Rev. Lett.} \textbf{\bibinfo{volume}{98}},
  \bibinfo{eid}{257204} (pages~\bibinfo{numpages}{4}) (\bibinfo{year}{2007}).

\bibitem[{\citenamefont{M\"uller and D\"oring}(1991)}]{Muller91ACIEE}
\bibinfo{author}{\bibfnamefont{A.}~\bibnamefont{M\"uller}} \bibnamefont{and}
  \bibinfo{author}{\bibfnamefont{J.}~\bibnamefont{D\"oring}},
  \bibinfo{journal}{Angew. Chem. Int. Ed. Engl.} \textbf{\bibinfo{volume}{27}},
  \bibinfo{pages}{157} (\bibinfo{year}{1991}).

\bibitem{footnote2} One may expect that if the applied field is high
  enough the spins align along its direction.

\bibitem[{\citenamefont{Procissi et~al.}(2006)\citenamefont{Procissi,
  Lascialfari, Micotti, Bertassi, Carretta, Furukawa, and
  K\"ogerler}}]{Procissi06PRB}
\bibinfo{author}{\bibfnamefont{D.}~\bibnamefont{Procissi}},
  \bibinfo{author}{\bibfnamefont{A.}~\bibnamefont{Lascialfari}},
  \bibinfo{author}{\bibfnamefont{E.}~\bibnamefont{Micotti}},
  \bibinfo{author}{\bibfnamefont{M.}~\bibnamefont{Bertassi}},
  \bibinfo{author}{\bibfnamefont{P.}~\bibnamefont{Carretta}},
  \bibinfo{author}{\bibfnamefont{Y.}~\bibnamefont{Furukawa}}, \bibnamefont{and}
  \bibinfo{author}{\bibfnamefont{P.}~\bibnamefont{K\"ogerler}},
  \bibinfo{journal}{Phys. Rev. B} \textbf{\bibinfo{volume}{73}},
  \bibinfo{pages}{184417} (\bibinfo{year}{2006}).

\bibitem[{\citenamefont{Chiorescu et~al.}(2000)\citenamefont{Chiorescu, Giraud,
  Jansen, Caneschi, and Barbara}}]{Chiorescu00PRL}
\bibinfo{author}{\bibfnamefont{I.}~\bibnamefont{Chiorescu}},
  \bibinfo{author}{\bibfnamefont{R.}~\bibnamefont{Giraud}},
  \bibinfo{author}{\bibfnamefont{A.~G.~M.} \bibnamefont{Jansen}},
  \bibinfo{author}{\bibfnamefont{A.}~\bibnamefont{Caneschi}}, \bibnamefont{and}
  \bibinfo{author}{\bibfnamefont{B.}~\bibnamefont{Barbara}},
  \bibinfo{journal}{Phys. Rev. Lett.} \textbf{\bibinfo{volume}{85}},
  \bibinfo{pages}{4807} (\bibinfo{year}{2000}).

\bibitem[{\citenamefont{Chiorescu et~al.}(2003)\citenamefont{Chiorescu,
  Wernsdorfer, M\"uller, Miyashita, and Barbara}}]{Chiorescu03PRB}
\bibinfo{author}{\bibfnamefont{I.}~\bibnamefont{Chiorescu}},
  \bibinfo{author}{\bibfnamefont{W.}~\bibnamefont{Wernsdorfer}},
  \bibinfo{author}{\bibfnamefont{A.}~\bibnamefont{M\"uller}},
  \bibinfo{author}{\bibfnamefont{S.}~\bibnamefont{Miyashita}},
  \bibnamefont{and} \bibinfo{author}{\bibfnamefont{B.}~\bibnamefont{Barbara}},
  \bibinfo{journal}{Phys. Rev. B} \textbf{\bibinfo{volume}{67}},
  \bibinfo{pages}{020402(R)} (\bibinfo{year}{2003}).

\bibitem[{\citenamefont{Lee et~al.}(1998)\citenamefont{Lee, Kilcoyne, and
  Cywinski}}]{Kilcoyne98}
\bibinfo{author}{\bibfnamefont{S.~L.} \bibnamefont{Lee}},
  \bibinfo{author}{\bibfnamefont{S.~H.} \bibnamefont{Kilcoyne}},
  \bibnamefont{and} \bibinfo{author}{\bibfnamefont{R.}~\bibnamefont{Cywinski}},
  \emph{\bibinfo{title}{Muon Science}} (\bibinfo{publisher}{SUSSP and Institute
  of Physics Publishing}, \bibinfo{year}{1998}).

\bibitem[{\citenamefont{Chow et~al.}(1998)\citenamefont{Chow, Hitti, and
  Kiefl}}]{Chow98}
\bibinfo{author}{\bibfnamefont{K.~H.} \bibnamefont{Chow}},
  \bibinfo{author}{\bibfnamefont{B.}~\bibnamefont{Hitti}}, \bibnamefont{and}
  \bibinfo{author}{\bibfnamefont{R.~F.} \bibnamefont{Kiefl}}, in
  \emph{\bibinfo{booktitle}{Semiconductors and Semimetals}}, edited by
  \bibinfo{editor}{\bibfnamefont{M.}~\bibnamefont{Stavola}}
  (\bibinfo{publisher}{Academic Press, New York}, \bibinfo{year}{1998}), vol.
  \bibinfo{volume}{51A}, p. \bibinfo{pages}{137}.

\bibitem[{\citenamefont{Salman et~al.}(2002)\citenamefont{Salman, Keren,
  Mendels, Marvaud, Scuiller, Verdaguer, Lord, and Baines}}]{Salman02PRB}
\bibinfo{author}{\bibfnamefont{Z.}~\bibnamefont{Salman}},
  \bibinfo{author}{\bibfnamefont{A.}~\bibnamefont{Keren}},
  \bibinfo{author}{\bibfnamefont{P.}~\bibnamefont{Mendels}},
  \bibinfo{author}{\bibfnamefont{V.}~\bibnamefont{Marvaud}},
  \bibinfo{author}{\bibfnamefont{A.}~\bibnamefont{Scuiller}},
  \bibinfo{author}{\bibfnamefont{M.}~\bibnamefont{Verdaguer}},
  \bibinfo{author}{\bibfnamefont{J.~S.} \bibnamefont{Lord}}, \bibnamefont{and}
  \bibinfo{author}{\bibfnamefont{C.}~\bibnamefont{Baines}},
  \bibinfo{journal}{Phys. Rev. B} \textbf{\bibinfo{volume}{65}},
  \bibinfo{pages}{132403} (\bibinfo{year}{2002}).

\bibitem[{\citenamefont{Uemura et~al.}(1985)\citenamefont{Uemura, Yamazaki,
  Harshman, Senba, and Ansaldo}}]{Uemura85PRB}
\bibinfo{author}{\bibfnamefont{Y.~J.} \bibnamefont{Uemura}},
  \bibinfo{author}{\bibfnamefont{T.}~\bibnamefont{Yamazaki}},
  \bibinfo{author}{\bibfnamefont{D.~R.} \bibnamefont{Harshman}},
  \bibinfo{author}{\bibfnamefont{M.}~\bibnamefont{Senba}}, \bibnamefont{and}
  \bibinfo{author}{\bibfnamefont{E.~J.} \bibnamefont{Ansaldo}},
  \bibinfo{journal}{Phys. Rev. B} \textbf{\bibinfo{volume}{31}},
  \bibinfo{pages}{546} (\bibinfo{year}{1985}).

\bibitem{footnote1} This relaxation function does not contradict that
  used in Ref. \onlinecite{Procissi06PRB}, however, transverse field
  muon precession measurements show no evidence for three inequivalent
  muon sites in the sample. Therefore we use this function to estimate
  the relaxation rate averaged over all sites.

\bibitem[{\citenamefont{Schneider et~al.}(1993)\citenamefont{Schneider, Kiefl,
  Chow, Johnston, Sonier, Estle, Hitti, Lichti, Connell, Sellschop
  et~al.}}]{Schneider93PRL}
\bibinfo{author}{\bibfnamefont{J.~W.} \bibnamefont{Schneider}},
  \bibinfo{author}{\bibfnamefont{R.~F.} \bibnamefont{Kiefl}},
  \bibinfo{author}{\bibfnamefont{K.~H.} \bibnamefont{Chow}},
  \bibinfo{author}{\bibfnamefont{S.}~\bibnamefont{Johnston}},
  \bibinfo{author}{\bibfnamefont{J.}~\bibnamefont{Sonier}},
  \bibinfo{author}{\bibfnamefont{T.~L.} \bibnamefont{Estle}},
  \bibinfo{author}{\bibfnamefont{B.}~\bibnamefont{Hitti}},
  \bibinfo{author}{\bibfnamefont{R.~L.} \bibnamefont{Lichti}},
  \bibinfo{author}{\bibfnamefont{S.~H.} \bibnamefont{Connell}},
  \bibinfo{author}{\bibfnamefont{J.~P.~F.} \bibnamefont{Sellschop}},
  \bibnamefont{et~al.}, \bibinfo{journal}{Phys. Rev. Lett.}
  \textbf{\bibinfo{volume}{71}}, \bibinfo{pages}{557} (\bibinfo{year}{1993}).

\bibitem[{\citenamefont{Blundell and Pratt}(2004)}]{Blundell04JPCM}
\bibinfo{author}{\bibfnamefont{S.~J.} \bibnamefont{Blundell}} \bibnamefont{and}
  \bibinfo{author}{\bibfnamefont{F.~L.} \bibnamefont{Pratt}},
  \bibinfo{journal}{J. Phys. Cond. Mat.} \textbf{\bibinfo{volume}{16}},
  \bibinfo{pages}{R771} (\bibinfo{year}{2004}).

\bibitem[{\citenamefont{Lascialfari et~al.}(1998)\citenamefont{Lascialfari,
  Jang, Borsa, Carretta, and Gatteschi}}]{Lascialfari98PRL}
\bibinfo{author}{\bibfnamefont{A.}~\bibnamefont{Lascialfari}},
  \bibinfo{author}{\bibfnamefont{Z.~H.} \bibnamefont{Jang}},
  \bibinfo{author}{\bibfnamefont{F.}~\bibnamefont{Borsa}},
  \bibinfo{author}{\bibfnamefont{P.}~\bibnamefont{Carretta}}, \bibnamefont{and}
  \bibinfo{author}{\bibfnamefont{D.}~\bibnamefont{Gatteschi}},
  \bibinfo{journal}{Phys. Rev. Lett.} \textbf{\bibinfo{volume}{81}},
  \bibinfo{pages}{3773} (\bibinfo{year}{1998}).

\bibitem[{\citenamefont{Borsa et~al.}(2006)\citenamefont{Borsa, Lascialfari,
  and Furukawa}}]{Borsa06}
\bibinfo{author}{\bibfnamefont{F.}~\bibnamefont{Borsa}},
  \bibinfo{author}{\bibfnamefont{A.}~\bibnamefont{Lascialfari}},
  \bibnamefont{and} \bibinfo{author}{\bibfnamefont{Y.}~\bibnamefont{Furukawa}},
  in \emph{\bibinfo{booktitle}{Novel NMR and EPR Techniques}}, edited by
  \bibinfo{editor}{\bibfnamefont{J.}~\bibnamefont{Dolinsek}},
  \bibinfo{editor}{\bibfnamefont{M.}~\bibnamefont{Vilfan}}, \bibnamefont{and}
  \bibinfo{editor}{\bibfnamefont{S.}~\bibnamefont{Zumer}}
  (\bibinfo{publisher}{Springer Berlin / Heidelberg}, \bibinfo{year}{2006}),
  pp. \bibinfo{pages}{297--349}.

\bibitem[{\citenamefont{Borsa}(2007)}]{Borsa07}
\bibinfo{author}{\bibfnamefont{F.}~\bibnamefont{Borsa}}, in
  \emph{\bibinfo{booktitle}{{NMR-MRI, $\mu$SR and Mossbauer spectroscopies in
  molecular magnets}}}, edited by
  \bibinfo{editor}{\bibfnamefont{P.}~\bibnamefont{Carretta}} \bibnamefont{and}
  \bibinfo{editor}{\bibfnamefont{A.}~\bibnamefont{Lascialfari}}
  (\bibinfo{publisher}{Spinger-Verlag}, \bibinfo{year}{2007}).

\bibitem[{\citenamefont{Yoneda et~al.}(2003)\citenamefont{Yoneda, Goto, Fujii,
  Barbara, and M\"uller}}]{Yoneda03PB}
\bibinfo{author}{\bibfnamefont{H.}~\bibnamefont{Yoneda}},
  \bibinfo{author}{\bibfnamefont{T.}~\bibnamefont{Goto}},
  \bibinfo{author}{\bibfnamefont{Y.}~\bibnamefont{Fujii}},
  \bibinfo{author}{\bibfnamefont{B.}~\bibnamefont{Barbara}}, \bibnamefont{and}
  \bibinfo{author}{\bibfnamefont{A.}~\bibnamefont{M\"uller}},
  \bibinfo{journal}{Physica B} \textbf{\bibinfo{volume}{329-333}},
  \bibinfo{pages}{1176} (\bibinfo{year}{2003}).

\bibitem[{\citenamefont{Furukawa et~al.}(2005)\citenamefont{Furukawa,
  Fujiyoshi, Kumagai, and K\"{o}gerler}}]{Furukawa05P}
\bibinfo{author}{\bibfnamefont{Y.}~\bibnamefont{Furukawa}},
  \bibinfo{author}{\bibfnamefont{Y.}~\bibnamefont{Fujiyoshi}},
  \bibinfo{author}{\bibfnamefont{K.}~\bibnamefont{Kumagai}}, \bibnamefont{and}
  \bibinfo{author}{\bibfnamefont{P.}~\bibnamefont{K\"{o}gerler}},
  \bibinfo{journal}{Polyhedron} \textbf{\bibinfo{volume}{24}},
  \bibinfo{pages}{2737} (\bibinfo{year}{2005}).

\bibitem[{\citenamefont{Lowe and Tse}(1968)}]{Lowe68PR}
\bibinfo{author}{\bibfnamefont{I.~J.} \bibnamefont{Lowe}} \bibnamefont{and}
  \bibinfo{author}{\bibfnamefont{D.}~\bibnamefont{Tse}},
  \bibinfo{journal}{Phys. Rev.} \textbf{\bibinfo{volume}{166}},
  \bibinfo{pages}{279} (\bibinfo{year}{1968}).

\bibitem[{\citenamefont{Dunsiger et~al.}(2006)\citenamefont{Dunsiger, Kiefl,
  Chakhalian, Greedan, MacFarlane, Miller, Morris, Price, Raju, and
  Sonier}}]{Dunsiger06PRB}
\bibinfo{author}{\bibfnamefont{S.~R.} \bibnamefont{Dunsiger}},
  \bibinfo{author}{\bibfnamefont{R.~F.} \bibnamefont{Kiefl}},
  \bibinfo{author}{\bibfnamefont{J.~A.} \bibnamefont{Chakhalian}},
  \bibinfo{author}{\bibfnamefont{J.~E.} \bibnamefont{Greedan}},
  \bibinfo{author}{\bibfnamefont{W.~A.} \bibnamefont{MacFarlane}},
  \bibinfo{author}{\bibfnamefont{R.~I.} \bibnamefont{Miller}},
  \bibinfo{author}{\bibfnamefont{G.~D.} \bibnamefont{Morris}},
  \bibinfo{author}{\bibfnamefont{A.~N.} \bibnamefont{Price}},
  \bibinfo{author}{\bibfnamefont{N.~P.} \bibnamefont{Raju}}, \bibnamefont{and}
  \bibinfo{author}{\bibfnamefont{J.~E.} \bibnamefont{Sonier}},
  \bibinfo{journal}{Phys. Rev. B} \textbf{\bibinfo{volume}{73}},
  \bibinfo{pages}{172418} (\bibinfo{year}{2006}).

\bibitem[{\citenamefont{Salman et~al.}(2000)\citenamefont{Salman, Keren,
  Mendels, Scuiller, and Verdaguer}}]{Salman00PB}
\bibinfo{author}{\bibfnamefont{Z.}~\bibnamefont{Salman}},
  \bibinfo{author}{\bibfnamefont{A.}~\bibnamefont{Keren}},
  \bibinfo{author}{\bibfnamefont{P.}~\bibnamefont{Mendels}},
  \bibinfo{author}{\bibfnamefont{A.}~\bibnamefont{Scuiller}}, \bibnamefont{and}
  \bibinfo{author}{\bibfnamefont{M.}~\bibnamefont{Verdaguer}},
  \bibinfo{journal}{Physica B} \textbf{\bibinfo{volume}{289-290}},
  \bibinfo{pages}{106} (\bibinfo{year}{2000}).

\end{thebibliography}
\end{document}